\documentclass[11pt,titlepage]{amsart}

\usepackage[]{atmcs} % for anonymous submission
\usepackage{hyperref,caption}

\usepackage{todonotes}
\usepackage{comment}
\usepackage{lineno}
\begin{document}

\title[Detecting the Indian Monsoon]{Detecting the Indian Monsoon using \\ Topological Data Analysis}

\author[EA]{Enrique G Alvarado}
\address{Iowa State University\\
    Ames, Iowa, USA}
\email{enrique3@iastate.edu}

\author[DB]{Daniela Beckelhymer}
\address{University of Minnesota\\    
    Minneapolis, Minnesota, USA}
\email{beck1538@umn.edu}

\author[JD]{Joshua Dorrington}
\address{University of Bergen\\
    Bergen, Norway}
\email{joshua.dorrington@uib.no}

\author[TL]{Tung Lam}
\address{University at Albany, State University of New York\\
    Albany, New York, USA}
\email{tlam@albany.edu}

\author[SM]{Sushovan Majhi}
\address{George Washington University\\
    Washington, D.C., USA}
\email{s.majhi@gwu.edu}

\author[JN]{Jasmine Noory}
\address{University of Minnesota\\    
    Minneapolis, Minnesota, USA}
\email{noory003@umn.edu}

\author[MM]{María Sánchez Muniz}
\address{City College of New York\\
    New York, New York, USA}
\email{msanchezmuniz@ccny.cuny.edu}

\author[KS]{Kristian Strommen}
\address{University of Oxford\\
    Oxford, UK}
\email{kristian.strommen@physics.ox.ac.uk}


\begin{abstract}
A monsoon is a wind system that seasonally reverses its direction, accompanied by corresponding changes in precipitation. The Indian monsoon is the most prominent monsoon system, primarily affecting India’s rainy season and its surrounding lands and water bodies. Every year, the onset and withdrawal of this monsoon happens sometime in May-June and September-October, respectively. Since monsoons are very complex systems governed by various weather factors with random noise, the yearly variability in the dates is significant. Despite the best efforts by the India Meteorological Department (IMD) and the South Asia Climate Outlook Forum (SCOF), forecasting the exact dates of onset and withdrawal, even within a week, is still an elusive problem in climate science.

%The onset and withdrawal days can be attributed to sudden changes of regime (transition to and from chaos) in the weather. During such a transition to chaos, topological signatures such as the Persistence Diagram of the system change drastically within a short period. The dynamics of the Indian monsoon depend on many weather factors, such as rainfall, temperature differential, wind speed, etc. A good starting point for analyzing the weather dynamics is to use Takens’ embedding on the monsoon index to reconstruct the phase space. Then, we detect the transition to chaos using topological data analysis (TDA) by tracking the history of the death and birth of -dimensional topological features. Recently, TDA has successfully detected chaos and approximated bifurcation diagrams for known dynamical systems, like the Lorenz system.

We interpret the onset and withdrawal of the Indian monsoon as abrupt regime shifts into and out of chaos. During these transitions, topological signatures (e.g., persistence diagrams) show rapid fluctuations, indicative of chaotic behavior. To detect these shifts, we reconstruct the phase space using Takens’ embedding of the Indian monsoon index and apply topological data analysis (TDA) to track the birth and death of $k$-dimensional features. Applying this approach to historical monsoon index data (1948–2015) suggests a promising framework for more accurate detection of monsoon onset and withdrawal.

%We use the historical data (1948-2015) of the Indian monsoon index to develop an early warning system for the onset and withdrawal of the Indian monsoon. In addition to giving specific onset and withdrawal dates, the proposed warning system also produces statically significant confidence bands (widows of dates) to predict transitions for a given level of significance.
\end{abstract}

\maketitle

\section{Introduction}
\label{sec:intro}

\begin{figure}[htb]
\centering
\includegraphics[width=0.5\linewidth]{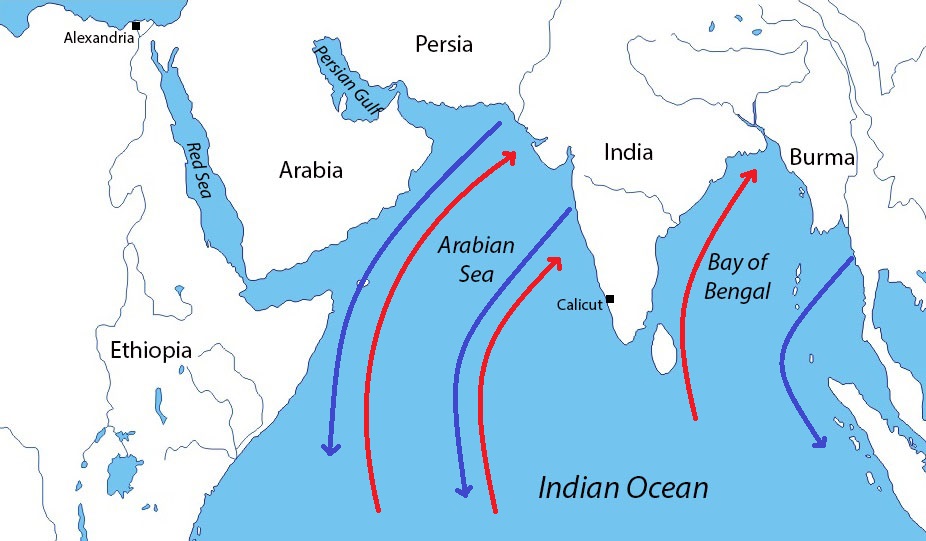}
\caption{The red and blue lines indicate the direction of wind over the Indian subcontinent during and after monsoon, respectively.}
\label{fig:wind}
\end{figure}

The Indian Monsoon is a seasonal reversing wind system that dramatically alters precipitation patterns over the Indian subcontinent, making it one of the most prominent and influential monsoon systems in the world \cite{gadgil2003indian}; see Figure~\ref{fig:wind}. Characterized by a clear seasonal reversal, the system is defined by southwest winds during the summer months that bring heavy rainfall and northeast winds during the winter that usher in drier conditions. This seasonal rhythm is vital for agriculture as the bulk of annual rainfall is delivered during the monsoon, ensuring water availability for crops.

\begin{comment}
    a robust monsoon can lead to abundant harvests, while a weak monsoon may trigger droughts and crop failures, thereby affecting food security and rural livelihoods.

Beyond its direct human and economic impacts, ranging from water resource management and electricity generation to overall economic stability and public health, the Indian monsoon serves as a natural laboratory for scientists(\cite{}).
\end{comment}  
Despite being driven by predictable seasonal changes, the monsoon exhibits dynamically chaotic behavior. Its transitions, particularly the onset and withdrawal phases, are marked by high sensitivity to initial atmospheric conditions and nonlinear dynamics---placing the system among the class of ubiquitous chaotic multi-state systems that remain poorly understood. Every year, the onset and withdrawal take place sometime in May--June and September--October, respectively. However, detecting the precise moments of monsoon onset and withdrawal is critical.

The Indian Monsoon stands as a bistable system, oscillating between distinct active and break phases \cite{gadgil2003indian}. 
The significance of these phase transitions extends beyond the subcontinent, influencing climate teleconnections worldwide, as demonstrated by Plumb et al. \cite{plum2021global}.
Early conceptual models from Webster et al. \cite{webster1998monsoons} captured the monsoon’s dual states, providing a theoretical basis for improved detection methodologies. 
Despite such foundational work, conventional approaches have sometimes struggled to pinpoint the subtle yet crucial shifts that mark the onset and withdrawal of the monsoon, underscoring the need for new analytical techniques.

Recent advances in \emph{Topological Data Analysis} (TDA) offer promising avenues for addressing this challenge by systematically extracting and quantifying topological invariants—connected components, loops, and higher-dimensional voids. A central tool in this framework, \emph{persistent homology}, tracks the birth, death, and lifespan of topological features under varying scales, providing a global perspective on regime changes and effectively capturing both subtle and abrupt transitions. Indeed, recent work \cite{strommen2023topological} shows that complex multistate systems often exhibit nontrivial topological characteristics, suggesting deeper connections between topology and nonlinear dynamical behavior. Moreover, when TDA is integrated with dynamical systems theory, it becomes possible to identify topological structures that might otherwise be masked by noise or incomplete data \cite{climatescienceTDA}. Accordingly, TDA is particularly well-suited for investigating monsoon onset, where complex, nonlinear interactions give rise to subtle dynamical transitions that traditional methods often fail to detect. 

Our approach applies TDA to monsoon studies by focusing on the topological signatures that highlight the Indian Monsoon’s inherent bimodality. 
By capturing critical changes via the persistence landscape, we demonstrate clear transitions between active and break phases. 
In particular, our topological landscape method leverages persistent homology to track these shifts, building on the dual-state framework originally established by Webster et al. \cite{webster1998monsoons}.

\begin{figure}[hbt]
    \centering
    \includegraphics[scale=.5]{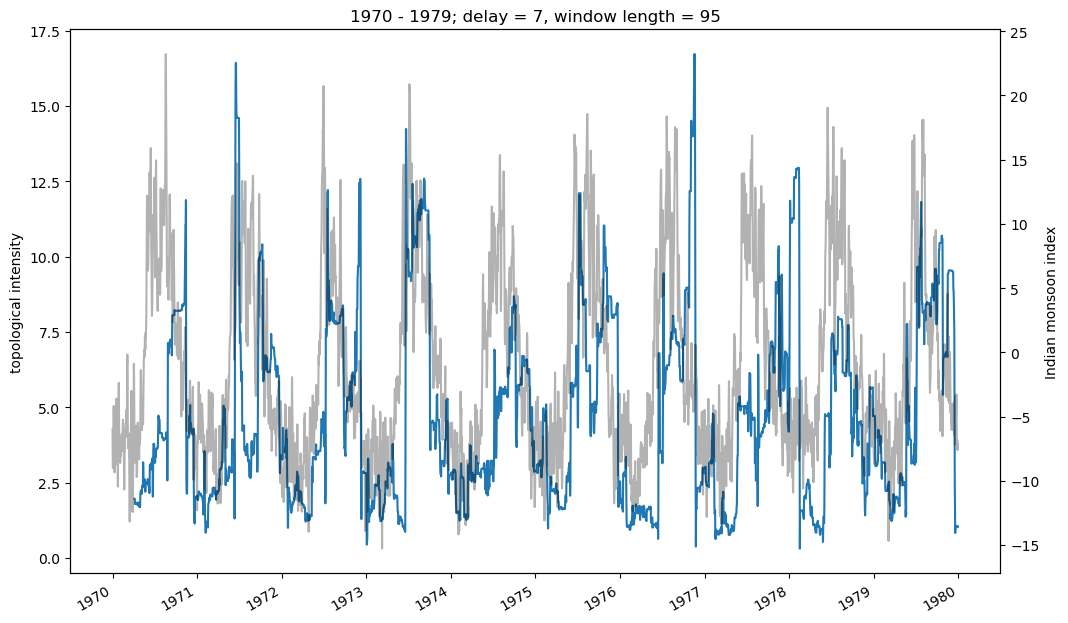}
    \caption{\small The monsoon index time series for the period 1970–1979 is shown in gray, while the time-series of the $\Lambda_{7, 95}$ is plotted in blue.}
    \label{fig:1970s}
\end{figure}

\section{Persistent Homology for Monsoon Onset}
%TDA provides a data-driven framework for systematically extracting and quantifying topological invariants—such as connected components, loops, and higher-dimensional voids—from complex datasets using methods like \emph{persistent homology}. 
%By examining the stability of these features across multiple scales, TDA offers a robust means of identifying subtle topological signatures. 
%This capability is especially pertinent in the study of multistate systems, where phase transitions may otherwise remain undetected. 
%Recent work \cite{strommen2023topological} has demonstrated that such systems frequently exhibit nontrivial topological characteristics, suggesting a deeper interplay between topology and dynamic behavior. 
%In climate research, combining TDA with dynamical systems theory has proven particularly valuable, as it facilitates the identification of persistent structures within high-dimensional models and time series—structures that could otherwise be masked by noise or limited observational windows \cite{climatescienceTDA}. 
%By tracking the birth, death, and lifespan of topological features under varying parameters, TDA yields a global perspective on regime changes, capturing both subtle and abrupt transitions. 
%Accordingly, TDA is particularly well-suited for investigating monsoon onset, where complex, nonlinear interactions give rise to subtle dynamical transitions that can be challenging to detect with traditional methods.
\subsection{Time-series analysis for Monsoon index}

The daily monsoon index data is given as a (discrete) time series $f: \{1,\ldots, N \} \to \R$, where each $f(i)$ represents the \emph{Indian Summer Monsoon index} on day $i$ from 1948 to 2015 \cite{apdrc}.
Inspired by Takens' time-delay embedding~\cite{takens2006detecting, Perea:amsnotice2019} and the use of persistence landscapes~\cite{JMLR:v16:bubenik15a}, we construct a time-dependent topological summary $\Lambda$ of the data and compare it with the original time series $f$.

\textbf{Time-Delay Embedding.}
We first embed the time series into a higher-dimensional space as follows: 
Given a \emph{delay} $d \in \N$, Takens' embedding maps $f$ into a sequence $\hat{f} = \{\hat{f}(i)\}_{i = d}^N$ in $\mathbb{R}^d$, where
\[
\hat{f}(i) = \bigl(f(i - d + 1), f(i - d + 2), \dots, f(i)\bigr) \quad \text{ for each } i = d, \dots, N.
\]
Note that each $\hat{f}(i)$ pertains to the monsoon index of $d$ consecutive days.
Takens' embedding theorem ensures that, once the attractor dimension of a dynamical system is known, one can select an appropriate delay so that the time-delay embedding is diffeomorphic to the original attractor~\cite{sauer1991embedology}. 
While the true attractor dimension is unknown in our setting, we find $d = 7$ to work well empirically. 
In future research, we will explore more systematic methods for determining $d$.

\textbf{Sliding Windows and Persistent Homology.}
%We let $w=90$ denote the length of our sliding window. For each index $i$ from $w$ to $N$, we define
%\[
%W_i \;:=\; \{\,i - w + 1,\, \dots,\, i\}
%\quad\text{and}\quad
%X_i \;=\; \{\hat{f}(j) : j \in W_i\} \;\subset\; \R^d.
%\]
%Since monsoon periods often span about 90 days, windows of length $w=90$ are natural for capturing season-scale trends.
%Hence, $X_i$ contains 90 consecutive (embedded) points. 
For any fixed {\it window length} $w \in \mathbb{N}$ and index $k$ from $w$ to $N$, we define the $k$-th {\it window} of $\hat{f}$ %$\{\hat{f}(i)\}_{i = 1}^N$ 
as  
\[
X_k \;:=\; \{\hat{f}(i) : i = k - w + 1,\, \dots,\, k\}.
\]
Here, each $X_k$ contains $w$ consecutive (embedded) points. 
We then construct the Vietoris--Rips complex $\mathrm{VR}_{\bullet}(X_k)$ and compute its one-dimensional persistent homology $H_1(\mathrm{VR}_{\bullet}(X_k))$. 
Each resulting persistence diagram, which tracks the birth and death of loops, is converted into a persistence landscape. 
We extract the first two landscape functions $\lambda_1^k, \lambda_2^k$ and compute their $L^2$-norm, yielding our main scalar of interest: 
\[
\Lambda_{d, w}(k) = (\|\lambda_1^k\|_2^2 + \|\lambda_2^k\|_2^2)^{1/2},
\]
representing the ``intensity'' of loop-like structures in the $k$-th window of $\hat{f}$.

%More, one can choose $w=30$, to captures intra-month variability while remaining short enough to highlight local structures.

\begin{figure}[h!]
    \centering
    \includegraphics[scale=0.5]{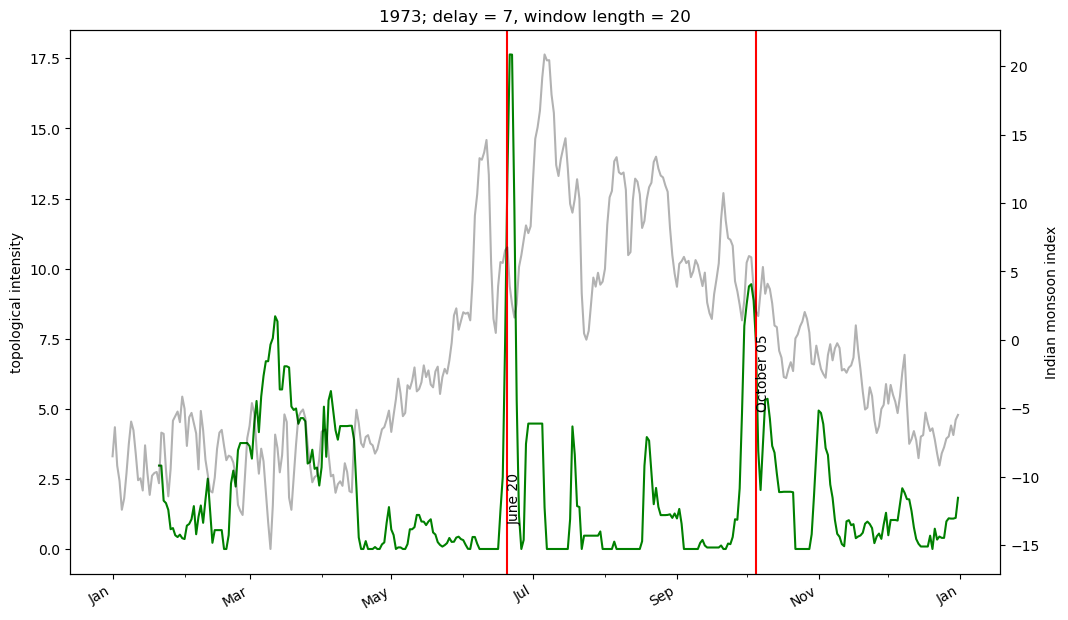}
    \caption{\small The monsoon index time series for the year 1973 is shown in gray, while the time-series of $\Lambda_{7, 20}$ are shown in green. The first red vertical line indicates the date when $\Lambda_{7, 20}$ exceeds $\mu+3\sigma$ for the first time; this is the predicted onset. The second red line indicates the date when the norm exceeds $\mu+2\sigma$ for the last time; this is the predicted withdrawal.
    Here, $\sigma$ is the empirical standard deviation of $f$ in 1973. 
    The official monsoon onset is June 21st, and the official monsoon withdrawal date is September 27th \cite{dhm2022}.}
    \label{fig:norms-1972}
\end{figure}
\textbf{Signal Similarity and Detecting Abrupt Changes.}
We now discuss relationships that we found between our topological signal $\Lambda_{d, w}$ (with delay $d = 7$) and the monsoon index $f$ with the following two choices of window lengths. 
\begin{enumerate}
\item {\bf ($w = 95$).}
We claim that it is natural to choose a window length $w$ where, for each year, there will be a single window that contains exactly the points that pertain only to monsoon days.
However, since the duration of the monsoon changes from year to year, to pick a single $w$, we choose $w$ to be the median monsoon duration.
In the 1970s, the median monsoon duration was 95 days. 
Plotting $\Lambda_{7, 95}$ with $f$, we can see the similarities between the signals (see Figure~\ref{fig:1970s}).
\item {\bf ($w = 20$).}
Choosing a window length of $w=20$, to capture intra-month variability while remaining short enough to highlight local structures.
Plotting this time series highlights intervals with significant topological changes. 
Notably, sharp spikes often signal the onset or withdrawal of the monsoon, aligning with pronounced transitions in the underlying dynamics. 
See Figure~\ref{fig:norms-1972} for a representative visualization from 1972.
\end{enumerate}

%\section{Discussion}

\clearpage
\bibliography{references}{}

\begin{thebibliography}{10}

\bibitem{JMLR:v16:bubenik15a}
Peter Bubenik.
\newblock Statistical topological data analysis using persistence landscapes.
\newblock {\em Journal of Machine Learning Research}, 16(3):77--102, 2015.

\bibitem{climatescienceTDA}
Davide Faranda, Th{\'e}o Lacombe, Nina Otter, and Kristian Strommen.
\newblock Climate science at the interface between topological data analysis and dynamical systems theory.
\newblock {\em Notices of the American Mathematical Society}, 71:267--271., 2024.

\bibitem{gadgil2003indian}
Sulochana Gadgil.
\newblock The indian monsoon and its variability.
\newblock {\em Annual Review of Earth and Planetary Sciences}, 31(1):429--467, 2003.

\bibitem{apdrc}
Yoshiyuki Kajikawa and Bin Wang.
\newblock Monsoon monitoring.
\newblock \url{https://apdrc.soest.hawaii.edu/projects/monsoon/realtime-monidx.html}.
\newblock Accessed: 2025-03-14.

\bibitem{dhm2022}
Department of~Hydrology and Nepal Meteorology.
\newblock Monsoon onset and withdrawal in {N}epal.
\newblock \url{https://www.dhm.gov.np/uploads/dhm/climateService/monsoon_onset_n_withdrawal_English_6_June_20221.pdf}, 2022.
\newblock Accessed: 2025-03-14.

\bibitem{Perea:amsnotice2019}
Jose~A Perea.
\newblock Topological time series analysis.
\newblock {\em Notices of the American Mathematical Society}, 66(5):686--694, 2019.

\bibitem{plum2021global}
Alan Plumb.
\newblock {\em The global circulation of the atmosphere, Chapter 9: Dynamical Constraints on Monsoon Circulations}.
\newblock Princeton University Press, 2021.

\bibitem{sauer1991embedology}
Tim Sauer, James~A Yorke, and Martin Casdagli.
\newblock Embedology.
\newblock {\em Journal of statistical Physics}, 65:579--616, 1991.

\bibitem{strommen2023topological}
Kristian Strommen, Matthew Chantry, Joshua Dorrington, and Nina Otter.
\newblock A topological perspective on weather regimes.
\newblock {\em Climate Dynamics}, 60(5):1415--1445, 2023.

\bibitem{takens2006detecting}
Floris Takens.
\newblock Detecting strange attractors in turbulence.
\newblock In {\em Dynamical Systems and Turbulence, Warwick 1980: proceedings of a symposium held at the University of Warwick 1979/80}, pages 366--381. Springer, 2006.

\bibitem{webster1998monsoons}
P.~J. Webster, V.~O. Maga{\~n}a, T.~N. Palmer, J.~Shukla, R.~A. Tomas, M.~Yanai, and T.~Yasunari.
\newblock Monsoons: Processes, predictability, and the prospects for prediction.
\newblock {\em Journal of Geophysical Research}, 103(C7):14451--14510, 1998.

\end{thebibliography}
\bibliographystyle{plain}
\end{document}